\documentclass[journal]{IEEEtran}

\usepackage{xcolor,soul,framed} 

\colorlet{shadecolor}{yellow}
\usepackage[pdftex]{graphicx}

\usepackage[cmex10]{amsmath}
\usepackage{amsmath}
\usepackage{array}
\usepackage{mdwmath}
\usepackage{mdwtab}
\usepackage{eqparbox}
\usepackage{url}
\usepackage[
singlelinecheck=false 
]{caption}
\usepackage{algorithm}
\usepackage{algorithmic}
\usepackage{multirow}
\usepackage{subcaption}
\usepackage{booktabs} 

\begin{document}
\title{Hiding Your Signals: A Security Analysis of PPG-based Biometric Authentication}

\author{\IEEEauthorblockN{Lin Li\IEEEauthorrefmark{1},
Chao Chen\IEEEauthorrefmark{2},
Lei Pan\IEEEauthorrefmark{3}, 
Yonghang Tai\IEEEauthorrefmark{4},
Jun Zhang\IEEEauthorrefmark{1}, and
Yang Xiang\IEEEauthorrefmark{1}}

\IEEEauthorblockA{\IEEEauthorrefmark{1}Swinburne University of Technology, Australia}

\IEEEauthorblockA{\IEEEauthorrefmark{2}RMIT University, Australia}

\IEEEauthorblockA{\IEEEauthorrefmark{3}Deakin University, Australia}

\IEEEauthorblockA{\IEEEauthorrefmark{4}Yunnan Normal University, China}

\thanks{This work has been submitted to the IEEE for possible publication. Copyright may be transferred without notice, after which this version may no longer be accessible.}
}



\maketitle

\begin{abstract}

Recently, physiological signal-based biometric systems have received wide attention. 
Unlike traditional biometric features, physiological signals can not be easily compromised (usually unobservable to human eyes). 
Photoplethysmography (PPG) signal is easy to measure, making it more attractive than many other physiological signals for biometric authentication. 
However, with the advent of remote PPG (rPPG), unobservability has been challenged when the attacker can remotely steal the rPPG signals by monitoring the victim's face, subsequently posing a threat to PPG-based biometrics.
In PPG-based biometric authentication, current attack approaches mandate the victim's PPG signal, making rPPG-based attacks neglected.
In this paper, we firstly analyze the security of PPG-based biometrics, including user authentication and communication protocols.
We evaluate the signal waveforms, heart rate and inter-pulse-interval information extracted by five rPPG methods, including four traditional optical computing methods (CHROM, POS, LGI, PCA) and one deep learning method (CL\_rPPG).
We conducted experiments on five datasets (PURE, UBFC\_rPPG, UBFC\_Phys, LGI\_PPGI, and COHFACE) to collect a comprehensive set of results.
Our empirical studies show that rPPG poses a serious threat to the authentication system. The success rate of the rPPG signal spoofing attack in the user authentication system reached 0.35. The bit hit rate is 0.6 in inter-pulse-interval-based security protocols.
Further, we propose an active defence strategy to hide the physiological signals of the face to resist the attack. It reduces the success rate of rPPG spoofing attacks in user authentication to 0.05. The bit hit rate was reduced to 0.5, which is at the level of a random guess. Our strategy effectively prevents the exposure of PPG signals to protect users' sensitive physiological data.

\end{abstract}


%
\IEEEpeerreviewmaketitle



\section{Introduction}
\label{intro}

Over the past decade, biometric systems have provided high authentication accuracy and user convenience, driving the widespread deployment. 
However, with the popularity of biometrics, researchers have found that traditional biometric authentication is vulnerable to spoofing attacks.
Traditional biometric features, such as face and fingerprint recognition, can be directly observed with the human eye, an attacker can compromise the authentication system without strong technical knowledge.
Researchers are turning to a potentially reliable and unobservable biological feature, the physiological signal.

Common physiological signals, including Electrocardiogram, Electroencephalograph, and Photoplethysmography (PPG) are widely used as biometric features~\cite{WANG2020107381,9130771Huang,Hwang9130730}.
Compared to other physiological signals, PPG signals are easier to obtain through a light source and a light sensor, which makes it more widely used in life. 
For example, popular wearable devices use PPG sensors for heath monitoring --- Apple and Samsung.
Therefore, PPG signals are considered to be more practical and attractive than other physiological signals in real-world applications \cite{Hwang9130730}.

Currently, known threats against PPG-based biometric authentication require the collection of the victim's PPG signal. This requires close contact with the victim, which significantly limits the possibility of an attack.
However, the emergence of remote PPG (rPPG) signals has challenged the unobservable property of PPG signals, which means that PPG-based biometric authentication will lose the advantage of unobservability.
rPPG attempts to capture PPG-like signals by monitoring subtle changes in the colour of a person's facial skin.
This is usually done by analyzing the colour change of the face area in the video.
The rPPG signal has been successfully applied to infer biometric signals, such as monitoring heart rate \cite{hu2021eta,dasari2021evaluation} and predicting blood pressure.

Although some studies explore the use of rPPG signals to attack inter-pulse interval (IPI)-based security protocols \cite{Seepers7892894,Calleja3, karimian2019attack},
they do not discuss viable defence strategies.
In addition, there is still a research gap between rPPG signals and PPG-based biometric authentication. 
In an existing rPPG method, its validity is assessed mainly by heart rate, but signal morphological characteristics are often neglected. 
In PPG-based biometric authentication, the morphological characteristics of the signal are the main contributors to the uniqueness of the user.
In addition, the rPPG signal is extracted from the human faces, while the PPG signal is mostly collected from the fingertips or wrist. 
Different skin tissues and distance from the heart can lead to differences in shape and phase between PPG signals.
Thus, it is crucial to explore the potential of rPPG-based attacks to evaluate the security of PPG-based biometrics, which is one of the aims of this work.

\begin{figure*}[!ht]
  \centering
  \includegraphics[width=1\linewidth]{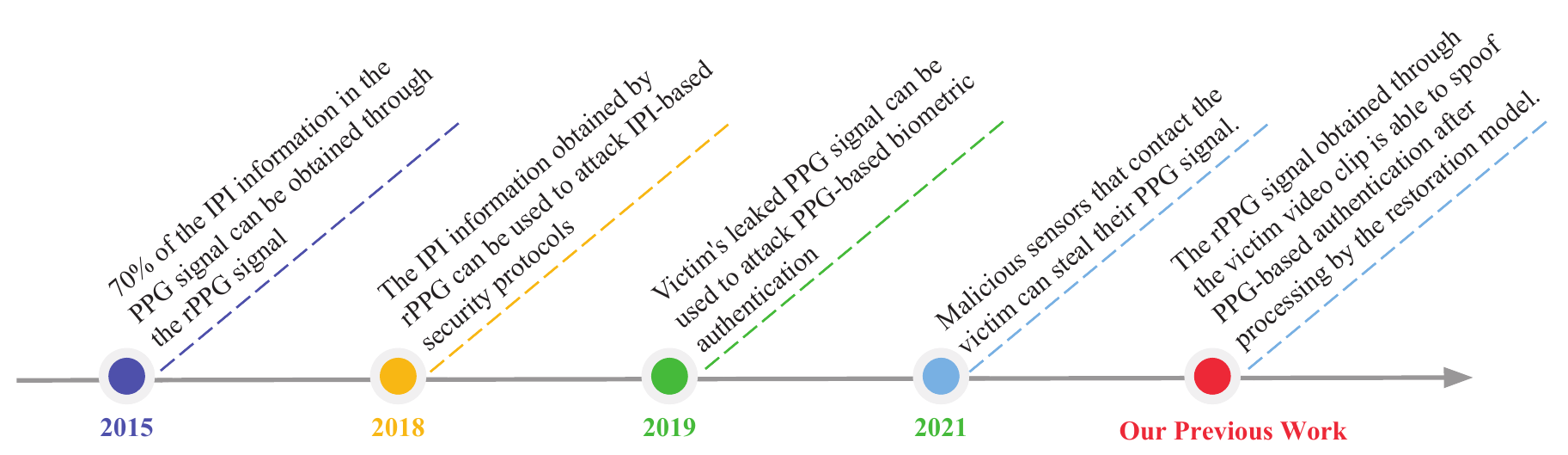}
  \caption{Evolution of the attack against PPG-based user authentication. The attack assumptions are released from requiring the victim's PPG signal to only needing the victim's video clip.}
  \label{timeline}
\end{figure*}

In this paper, we analyze the security of PPG-based biometric authentication, 
particularly in user authentication and communication protocols.
As shown in Fig.~\ref{timeline}, current methods only require access to video clips of the victim's face to achieve a remote attack, increasing the likelihood of the attack in reality.
We firstly analyze rPPG-based spoofing attacks and some common defence strategies to reduce the threat. 
We further propose an effective defence strategy and evaluate the defence performance.
The main contributions of this paper are as follows:

\begin{enumerate}
  \item We analyze the potential threats of rPPG to PPG-based biometrics using five methods (CHROM, POS, LGI, PCA, CL\_rPPG) on five datasets (PURE, UBFC\_rPPG, UBFC\_Phys, LGI\_PPGI, COHFACE).
  The success rate of the spoofing attack on user authentication reached 0.35. The bit hit rate in the IPI-based security protocols reached 0.6.

  \item We analyze a series of defence strategies to mitigate the threat of rPPG-based spoofing attacks on biometrics. Existing passive defense strategies of spoofing attacks can significantly degrade the performance of the user authentication model.

  \item We propose a signal hiding method (SigH) by injecting an arbitrary waveform into the face of the video. Our active defense strategy reduces the success rate of spoofing attacks on user authentication to 0.05. And it will not affect the performance of the authentication model. The bit hit rate in the IPI-based security protocols was reduced to 0.5.
  
\end{enumerate}

The rest of the paper is organized as follows: Section \ref{relatedwork}  reviews related work, including PPG-based biometric authentication methods, existing attacks, and rPPG methods.
Section~\ref{threatmodel} and~\ref{defence} describe the attack and defence strategies we used.
In Section~\ref{Attackresult} and~\ref{Defenceresult}, experimental results are provided along with a discussion on the performance of the attack and defense strategy, respectively.
Finally, Section \ref{conclusion} concludes this work.

\section{Related Work}
\label{relatedwork}

\subsection{PPG-based Biometrics}
In the biometrics field, PPG signals are mainly used in two areas: user authentication and communication protocols.

\begin{figure}[!ht]
  \centering
  \includegraphics[width=\linewidth ]{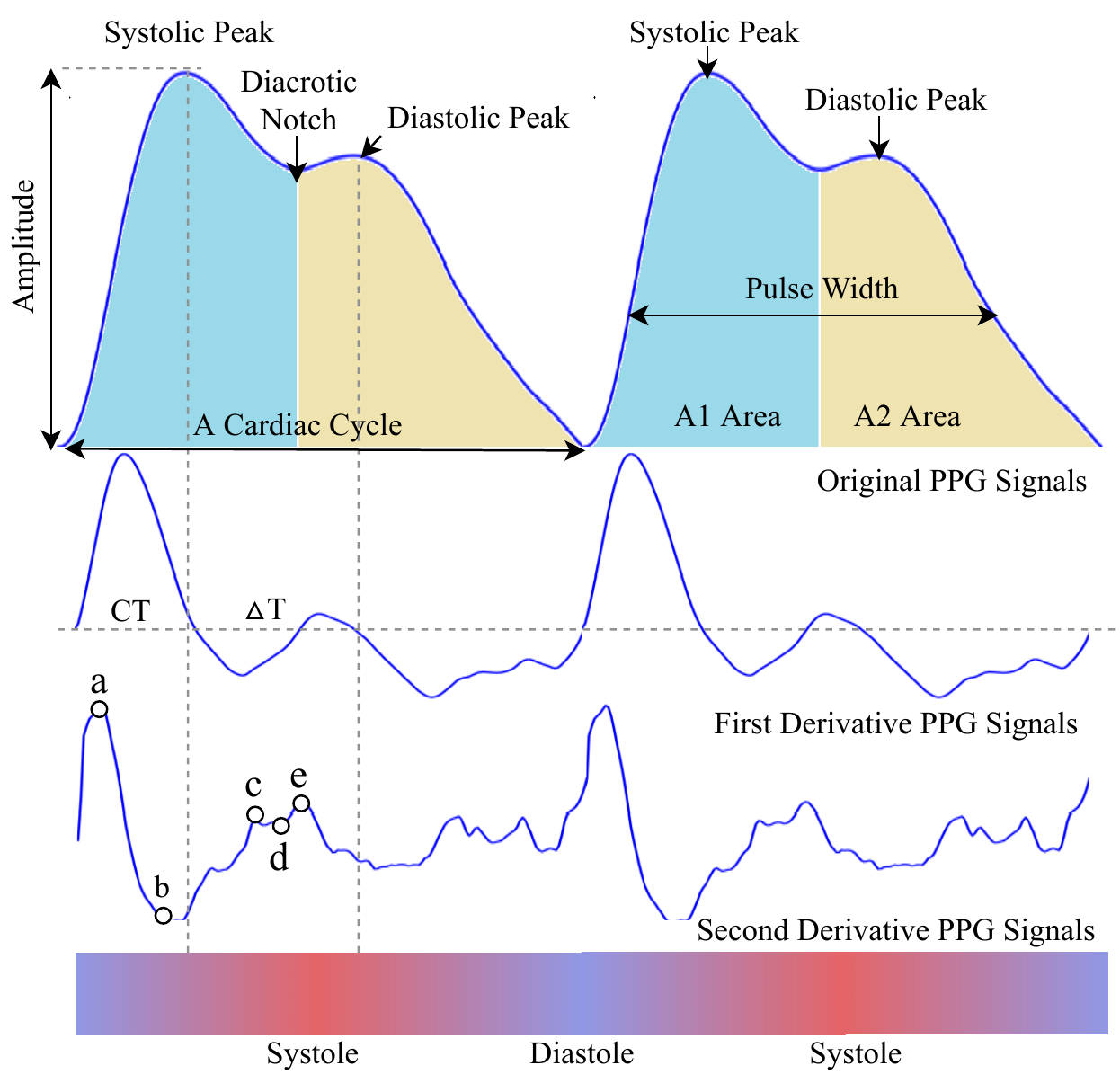}
  \caption{Features of the PPG signal (including the original waveform, the first-/second- order derivatives of the waveform) that are related to heart health. }
  \label{wave}
\end{figure}

\textbf{User Authentication:} 
The PPG signal contains a wealth of personal cardiac data.
The earliest studies used clinical features commonly existed in PPG signals as biometric features ~\cite{Gu1222403,RESITKAVSAOGLU20141}. For example, as shown in Fig.~\ref{wave}, the ratio of $b$ to $a$ is related to arterial stiffness \cite{takazawa1998assessment}. $A1$ and $A2$ areas are related to systemic vascular resistance \cite{awad2007relationship}. 
However, some characteristics of these features are sensitive to changes over time. For example, $\Delta T$ changes with an individual's age \cite{millasseau2002determination}. As a result, researchers have applied a few techniques to get stable features, such as filtering noise-sensitive features using feature selection methods like Kernel Principal Component Analysis \cite{Zhang3297174}, transforming robust features by leveraging wavelet transform \cite{Yadav8411233} and Short-time Fourier Transform \cite{donida2021biometric}. GAN has also been used to synthesize temporally stable features of individual human \cite{hwang2021pbgan}.

After constituting an individual template using the extracted signal features, the process of identifying the individual template is similar to other biometric authentication systems. 
The early generations of authentication systems match individuals by calculating the distance between features and templates \cite{akhter2015heart, Gu1222403}.
Gradually, machine learning models replace simple distance calculations to improve the performance of template matching.
However, many machine learning models require manual feature extraction so that the system designer and engineers may introduce bias.
Recently, deep learning methods have provided an end-to-end authentication strategy, for example, CNN and LSTM are used to learn features from raw data automatically in the context of user recognition \cite{Biswas8607019, Hwang9130730}.

\textbf{Communication Protocols:}  In body area sensor networks (BASN) with strict computational resource constraints, traditional encryption schemes for information transmission are difficult to apply because PPG signals can get similar measurements in different parts of the body. 
When used as an entity identifier in a BASN symmetric encryption scheme, the distribution of pre key-agreement could be avoided \cite{zhang2021h2k}. 
IPI information extracted from the PPG signal is the most prevalent entity identifier. 
IPI indicates the time difference between two consecutive heartbeats, while PPG signal is usually regarded as the time difference between two pulse peaks. 
The timing of systolic peaks in the PPG signal is sympathetically related and influenced by various physiological factors.
This characteristic allows the researcher to extract entity identifiers from the IPI information \cite{chizari2019extracting}.

After obtaining the IPI, the randomness of the entity identifier needs to be extracted by quantization algorithms. 
A quantization algorithm encodes the signal into a binary representation.
Two quantization methods exist. 
The first method was proposed in \cite{lin2019h2b} to extract a random value from the intermediate bits of IPI. 
However, this method suffers from obvious issues, that is, the high bits of IPI are not random enough while the low bits have too much noise. 
The second method was proposed in \cite{chizari2019extracting} to use the trend of the IPI sequence as a random source of information. 
The second method improves the randomness attribute at the cost of excessive processing time.

\subsection{Remote Photoplethysmography (rPPG)}
\label{Remoteppg}

Most rPPG signals are obtained by analyzing subtle differences in the facial colour channels from a video clip.
The green channel is more readily absorbed by hemoglobin than red and blue channels. 
In most circumstances, the green channel contains the strongest PPG signal, providing a high signal-to-noise ratio solution for signal acquisition \cite{verkruysse2008remote}.
Two blind source separation methods, Independent Component Analysis and Principal Component Analysis were prposed in \cite{poh2010non,lewandowska2011measuring} to isolate the approximately clean signals from the RGB channels.
Unlike contact PPG measurements, the signal reflected by the skin in rPPG measurements includes the specular reflection of natural light, leading to unpredictable normalization errors.
CHROM~\cite{de2013robust} introduces chromatic aberration to eliminate specular reflection effects with the assumption that the pixels in the face region contribute equally to the rPPG signal.
Nevertheless, pixels collected from the face region may be affected by different levels of noise.

A recent research trend is to apply deep learning in rPPG acquisition for an end-to-end solution framework.
DeepPhys \cite{chen2018deepphys} uses an attention network with CNN to capture the differences in the spatial and temporal distribution of signals between video frames.
PhysNet \cite{yu2019remote} uses a spatio-temporal convolutional network to restore the rPPG signal from the video clip.
Meta-rPPG \cite{lee2020meta} presents a transduction meta-learner that allows the deep learning model to adapt the data distribution during deployment.
A fully self-supervised training method was introduced in \cite{Gideon_2021_ICCV} to acquire the rPPG signals.
In addition to existing progress, deep learning-based methods have good research potentials for rPPG signal acquisition.

In summary, the research trend for rPPG signal acquisition is towards automated end-to-end solutions with high accuracy and resilience to noise. 
However, many existing research works are evaluated by the extracted heart rate's accuracy, neglecting the signal's morphological characteristics.

\subsection{Existing Attacks}

The first attack against IPI-based authentication using rPPG was proposed by Calleja et al.~\cite{Calleja3} in 2015. 
IPI signals help derive secure random sequences in the key distribution component of IPI-based authentication systems whose security conditions rely on the assumption that IPI can be measured by contact devices only.
In fact, most of the IPI information acquired by PPG signals could also be obtained by non-contact techniques (rPPG).
By analyzing the face's colour change information in the video clip, the rPPG signal can be extracted in a manner similar to PPG.
The extracted rPPG signal can be used to generate IPI, violating the assumptions in IPI-based authentication protocols \cite{Seepers7892894}.
Although rPPG does not accurately describe IPI generated from ECG signals, it has similar accuracy to contact PPG.
Currently, PPG-based biometric authentication is mainly based on the morphological characteristics of the PPG signal rather than the IPI information.

In previous attacks on PPG-based biometric authentication, researchers almost always assume that the victim's PPG signal has been compromised.
Karimian et al.~\cite{karimian2019attack} utilized a non-linear dynamic model \cite{mcsharry2003dynamical} to extract model parameters from the victim's PPG signal before transforming the model parameters using two Gaussian functions as mapping functions. 
Thereafter, a spoofing attack was launched using the forged victim's PPG signal.
Another study involved stealing a victim's PPG signal stealthily by installing a malicious PPG sensor on a device that the victim would touch \cite{hinatsu2021basic}. 
Based on the recorded signals, the attacker can use a waveform generator to simulate the victim's PPG signal and compromise the authentication.
However, contact-based PPG signals are challenging to obtain without the victim's awareness, significantly limiting these attacks' application scenarios.

\begin{figure*}[!ht]
  \centering
  \includegraphics[width=\linewidth]{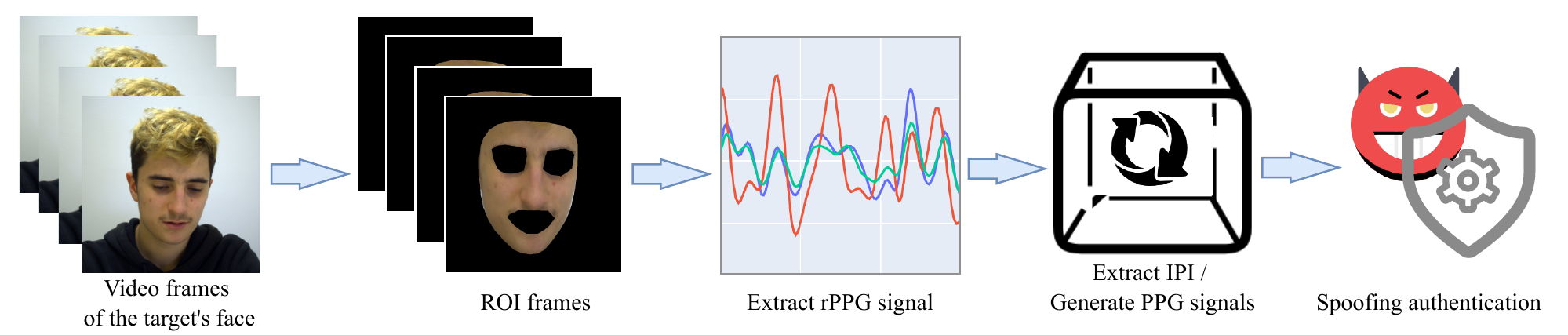}
  \caption{Our proposed spoofing attack flow. We identify the ROI region from the target video frame before extracting the rPPG signal from the ROI. The rPPG signal is used to launch a spoofing attack on the authentication system. Usually, we can use a waveform generator to convert the PPG signal to an electrical signal for transmission to the target terminal or simulate the process of PPG signal acquisition. For the IPI-based protocol, we can extract the IPI sequence from the rPPG signal.}
  \label{workflow}
\end{figure*}

\section{Spoofing attacks on PPG-based Biometric Authentication}
\label{threatmodel}

In this work, our goal is to conduct a comprehensive security analysis of PPG-based biometrics.
First, Fig.~\ref{workflow} depicts our scheme that performs spoofing attacks on biometrics.
We extracted the rPPG signal from the face video clip, then extracted the IPI and reconstructed the PPG signal from the rPPG signal. 
We consider various rPPG methods (CHROM, POS, LGI, PCA, CL\_rPPG) to spoof biometric authentication, as well as extracting IPI information from the rPPG signal to spoof IPI-based security protocols.

\subsection{Threat Model}
\label{assumption}

PPG-based biometrics, including user authentication and IPI-based security protocols, are vulnerable to spoofing attacks. 
This paper assumes that the attacker could obtain a video clip with the victim's face.
In this age of social networking, it is easy to retrieve many videos with human faces on social media sites.
For instance, YouTube and Facebook (Meta) host numerous HD video clips that are downloadable to the attacker.
Furthermore, an attacker can record a video clip of a targeted victim's face through a surveillance camera, especially when hidden cameras are often not perceived by the victim.
In contrast to the previous attack assumptions, our method requires no leaked PPG signals.
In addition, we assume that the attacker knows the inputs to the user authentication model. 
For instance, a PPG signal with one heartbeat cycle is taken as input. The length of the signal is resampled to 60. The amplitude of the signal is normalized to 0-1.

\subsection{rPPG Acquisition}
\label{signalacquisition}

Before acquiring the rPPG signal, we use MTCNN \cite{zhang2016joint} to detect the face region in each video frame before isolation.
Then the non-skin part of the face region is filtered by the skin detection algorithm proposed in \cite{kolkur2016human}.
For example, clothes, hair, and other parts that do not provide PPG signal information will be filterd. 
We will get a sequence of video frames $F(n): 0 < n < total\ number\ of\   frames$ containing only the facial skin. $n$ denotes the $n$th frame in the video clip. 
Each frame $F(n)$ is stored as a matrix of $w \times h \times 3$, where $w$ and $h$ represent the frame's width and height, and 3 is the number of color channels. 
The average value of RGB of the whole skin area is used as the R, G, B values of the current frame. 
We apply a bandpass filter to obtain the portion of the frame sequence with signal frequencies between 0.65Hz and 4.0Hz (i.e., between 39 and 240 bpm). 
The method in \cite{tarvainen2002advanced} was used to remove the breadth component of the signal. 
This preprocessing removes the signal's noise and breathing components (signal trend).

Upon removing noises, we extract the rPPG signal from the consecutive skin frames.
As listed in Section~\ref{Remoteppg}, there are multiple methods for obtaining rPPG signals. For instance, CHROM, POS, LGI, PCA, CL\_rPPG.

\textbf{CHROM} is based on the skin optical reflection model.
The model assumes that the light reflected by the skin consists of diffuse and specular reflections, and the PPG signal is hidden in the diffuse reflection. 
CHROM eliminates the specular reflection component by using chrominance.
CHROM uses simple mathematical operations to obtain the PPG signal quickly. 
Moreover, CHROM is resilient to motion artifacts.
$X$ and $Y$ represent two orthogonal axes, and $R$, $G$, and $B$ represent the three color channels in the video. 
The rPPG signal value $S$ can be calculated as follows:
\begin{equation}
  X = 3 \times R - 2 \times G 
\end{equation}
\begin{equation}
  Y = (1.5\times R)+G-(1.5 \times B)
\end{equation}
\begin{equation}
  \alpha =\sigma (X)/\sigma (Y)
\end{equation}
\begin{equation}
  S = X - \alpha \times Y
\end{equation}
\noindent where $\sigma$ represents the method of obtaining the standard deviation.

\textbf{POS} is similar to CHROM, which is also based on the skin reflection model assumption to obtain low signal-to-noise ratio PPG signals by removing specular reflections. 
First, POS uses a sliding window to normalize the RGB pixel values within the window temporally. 
$C(t) = [R(t), G(t), B(t)]^T$. Time normalization: 
\begin{equation}
 C_w(t) = \sum_{i=t}^{w}C(i)
\end{equation}
\begin{equation}
 C_{tn}(t) = \frac{ C_w(t)}{\mu ( C_w(t))}
\end{equation}

\noindent $w$ denotes the time window, and $t$ indicates the $t$-th frame.
The spatial pixel averaging step can reduce camera quantization errors. 
The signal is then projected into the plane $P$. 
Finally, we get the rPPG signal $S$. 

\begin{equation}
P = \begin{pmatrix}0 & 1&-1\\ -2 & 1&1\end{pmatrix}
\end{equation}
\begin{equation}
X, Y = P \cdot C_{tn}(t)
\end{equation}
\begin{equation}
  S = X + \alpha * Y
\end{equation}

\textbf{LGI} extracts the Local Group Invariance features of heart rates from face videos. 
The singular value decomposition of $C_{tn}(t)$ is first performed, such that $ C_{tn}(t) = U\Sigma V^{T}$.
The projection plane is defined as $P$:

\begin{equation}
P = I - UU^{T}
\end{equation}
\begin{equation}
S = P \cdot C_{tn}(t)
\end{equation}

Principal component analysis
(\textbf{PCA}) is a statistical analysis method to reduce features' dimensionality. 
PCA uses an orthogonal transformation to project the observed values of potentially correlated variables into a set of linearly uncorrelated variables. 
PCA separates the periodic pulse signals from the noisy signals.

\textbf{CL\_rPPG} is a deep learning-based rPPG signal extraction method. 
Traditional methods may lose vital information related to the heartbeat through manually designed features. 
The deep learning-based approach recovers the rPPG signal directly from the original face video. 
CL\_rPPG uses PhysNet as the rPPG estimator to generate the rPPG signal. 
PhysNet is a spatiotemporal convolutional network representing both long-term and short-term spatio-temporal contextual features. 
CL\_rPPG outputs the PPG signal through the upsampling interpolation and 3D convolution. 
CL\_rPPG has two versions: supervised learning (CL\_rPPG\_P) and self-supervised learning (CL\_rPPG\_F). 
For the supervised learning mode, the CL\_rPPG\_P model is trained using the maximum cross-correlation between the ground truth PPG signal and the estimated PPG signal as the loss function. 
In the self-supervised learning model CL\_rPPG\_F, negative samples with high heart rates are artificially generated by a frequency resampler for contrast training. 
The power spectral density means the squared error is used to measure the difference between two signals.

\subsection{IPI Recovery and Quantification}
\label{ipirecovery}
To spoof IPI-based security protocols, we need to recover the IPI information from the signal.
We mark the peak locations of the original signal before calculating the time difference between the peaks as IPI. 
As the standard heart rate in adults is 50-120 beats per minute \cite{hwang2021pbgan}, there are at most two heartbeats per second. 
To minimize sample loss, we exclude data where the distance between heartbeats is less than $FPS/2$.

For IPI sequences to be used in key exchange protocols, we need to quantize the IPI sequences. 
There are two mainstream quantification methods, trend-based and quantile function-based \cite{zhang2021h2k}.
We use the Gray code as the quantization function for the quantile-based function.
Firstly, we normalize the IPI time series to 0-1 by the following equation. 
\begin{equation}
IPI_s = \frac{IPI-IPI_{min}}{IPI_{max}-IPI_{min}}
\end{equation}
\noindent Then, we multiply it by 256 and convert the result to an 8-bit Gray code. 
\begin{equation}
IPI_{tmp} = IPI_s \times 256
\end{equation}
\noindent According to Equation \ref{togray}, the $bin()$ function converts decimal numbers to binary. The $zfill(8)$ function fills binary numbers with zeros up to eight bits.
\begin{equation}
\label{togray}
IPI_{G} = bin(IPI_{tmp}^{(IPI_{tmp}>>1)}).zfill(8)
\end{equation}

For the trend-based quantification, we divide each IPI value into 16 parts of an equal length following a normal distribution. Our setting is identical to IMDGuard \cite{xu2011imdguard}. 
Then we encode the IPI sequence according to Algorithm \ref{alg:IPI}.

\begin{algorithm}[!ht]
    \caption{IPI trend-based quantification}
    \label{alg:IPI}
    \begin{algorithmic}[1]
    \STATE $IPI_{T}=[0]$  \% Set the first position of the code to 0
    \STATE \% Sequential traversal of the IPI sequence trend
    \FOR {$i \ in \ IPI_s$}
        \IF {$i  \le  IPI_{T}[-1]$}
            \STATE $IPI_{T}.append(0)$ \% When trend decreases
        \ELSE
            \STATE $IPI_{T}.append(1)$ \%When trend rises
        \ENDIF
    \ENDFOR
    \end{algorithmic}
\end{algorithm}

\section{Attack Evaluation}
\label{Attackresult}

In this section, we evaluate the effectiveness of spoofing attacks using rPPG signals.
We select five common public datasets for our experiments.
Experiments on spoofing user authentication were performed in UBFC-PHYS \cite{Meziatisabour9346017}. We trained a state-of-the-art model as our user authentication target model (see Section~\ref{Implementation}). 
The threat of rPPG signals to IPI-based security protocols is explored on four datasets --- PURE \cite{stricker2014non}, UBFC\_rPPG \cite{bobbia2019unsupervised} (UBFC\_1, UBFC\_2),  LGI\_PPGI \cite{Pilz8575328} and COHFACE \cite{heusch2017reproducible}. We use trend-based and quantile function-based as our target model (see Section~\ref{ipirecovery}) for IPI-based security protocols. Then we use the rPPG signal (see Section~\ref{signalacquisition}) to perform a spoofing attack.

\subsection{Datasets}
\label{datasets}
To analyze the security of PPG-based authentication, we mainly consider the dataset, UBFC-PHYS \cite{Meziatisabour9346017}. 
Due to video quality, the rPPG signal recovered in other datasets is insufficient to threaten the biometrics.

\textbf{UBFC-PHYS} captures facial videos ($1024 \times 1024$ resolution, 35 FPS, 227,474 Kbps) of 56 participants. Participants are about 1 meter away from the camera and the light source. The light source uses artificial light to ensure the same lighting conditions.
Each participant recorded a one-minute video of the three states (`resting', `talking', or `calculating').
Meanwhile, the \textit{Empatica E4} wristband was used to collect the PPG signal from the wrist synchronously with the sampling rate of 65 Hz. 
We exclude some videos in which faces are not correctly recognized in the video frames due to the participants' activity.
We perform repeated independent experiments for each user. Other users are treated as non-victims.
As shown in Table \ref{testnumber}, we extracted rPPG signals ranging from 150 to 287 cardiac cycles for each state.
The number of cardiac cycles varies depending on the individual.

\begin{table}[!ht]
\centering
\begin{tabular}{|c|c|}
\hline
\textbf{Signals}& \textbf{Number of Cardiac Cycles} \\ \hline
Non-victim PPG Signal & 11800+   \\ \hline
Victim's rPPG Signal   & 150$\sim$287   \\ \hline
\end{tabular}
\caption{The cardiac cycles count for each victim's rPPG signal and non-victim PPG signal.}
\label{testnumber}
\end{table}

To analyze the threat of rPPG for IPI-based protocols, we considered the four most commonly used rPPG datasets, PURE \cite{stricker2014non}, UBFC\_rPPG \cite{bobbia2019unsupervised},  LGI\_PPGI \cite{Pilz8575328} and COHFACE \cite{heusch2017reproducible}.

\textbf{PURE} captures 10 participants' facial videos. 
Each participant's videos were recorded in six states (Steady, Talking, Small/Medium Rotation, and Slow/Fast Translation). 
Each video clip is one-minute long with a resolution of 640 $\times$ 480 and a frame rate of 30 FPS. 
Meanwhile, the finger clip pulse oximeter acquired the PPG signal at a 65Hz sampling rate. 
Although there was facial movement in the video, all signals were captured during the rest state.

\textbf{UBFC\_rPPG} uses a webcam to record video (640 $\times$ 480 resolution, 30 FPS). 
Transmissive pulse oximetry (62 Hz) to obtain the PPG signal. It consists of two sub-datasets. In UBFC\_1, participants were asked to remain stationary; in UBFC\_2, participants played a mathematical game in front of a green screen.

\textbf{LGI\_PPGI} records facial videos in four scenes (resting, head movement, fitness, and conversation in an urban background). The video clips were captured by a webcam (640 $\times$ 480 resolution, 25 FPS). The sampling rate of the pulse oximeter is 60 Hz. However, only 6 of the original 25 participants' videos are publicly available. Our experiments were conducted on these 6 participants.

\textbf{COHFACE} contains facial video clips (640 $\times$ 480 resolution, 20 FPS) of 40 subjects, including the simultaneous acquisition of PPG signals (256 Hz). Subjects were asked to sit still in front of the camera. The video was heavily compressed.

\begin{table}[!ht]
\renewcommand\arraystretch{1.2}
\centering
\setlength{\tabcolsep}{3mm}{\begin{tabular}{|l| c| c| c| }
\hline
\textbf{Status}  & \textbf{Resting}  & \textbf{Talking} & \textbf{Calculating}  \\\hline
Random Attack & 0.14   & 0.15   & 0.15  \\ \hline
Victim rPPG Signal Attack  & 0.25   & 0.19   & 0.21   \\ \hline
Mean rPPG Signal Attack & \textbf{0.34}   & \textbf{0.35}   & \textbf{0.35}   \\ \hline
\end{tabular}}
\caption{The success rate of spoofing attacks in UBFC-Phys. Resting, Talking, and Calculating represent the three states of the victim in the video. Mean indicates the average of the PPG signals over multiple cardiac cycles. The mean value is used by us to maximize the success rate of the attack victim.}
\label{result1}
\end{table}

\subsection{Experimental Implementation}
\label{Implementation}
We use Keras\footnote{\url{https://www.tensorflow.org/}} to implement the most advanced PPG-based user authentication model \cite{Hwang9130730}. 
We use the virtual heart rate pyVHR\footnote{\url{https://github.com/phuselab/pyVHR}} \cite{Boccignone9272290} to implement the different traditional rPPG methods (CHROM, POS, LGI, PCA). 
It provides a uniform standard of evaluation. 
For deep learning methods CL\_rPPG, we use the open-source code\footnote{\url{https://github.com/ToyotaResearchInstitute/RemotePPG/tree/master/iccv}} provided in the article.

For the user authentication model, we mark the signals of the victims as class 1 and other users as 0.
We aim to compare the threat of rPPG signals with other users' PPG signals for the user authentication system. 
We treat each user as a victim and one-tenth of other users are used to train the user authentication model, which is realistic for authentication systems. 
It is possible to collect a large amount of data about the target user, but only a small amount of data from other users. 
The remaining data from other users were used as a random attack to evaluate the spoofing attack by rPPG signals.
To explore the impact of signals collected in different states on spoofing attacks, we exclude the PPG signals in the `talking' and `calculating' states while training the user authentication model. 
The performance of the user authentication model is determined by the Equal Error Rate (EER). 
We also call the model for the false acceptance rate (FAR) of the rPPG signal as the success rate of spoofing. 
A higher success rate indicates that the authentication system is more vulnerable to such spoofing attacks. Finally, we use the average results of all users.

For IPI-based security protocols, we evaluate them by using heart rate (HR) and IPI, respectively. 
The primary metrics are mean absolute error (MAE) and root-mean-square deviation (RMSE).
We also use Pearson's coefficient (PC) to compare the similarity between the PPG signal and the rPPG signal. 
Since IPI-based security protocols require multi-bit encoding from IPI, we use the bit hit rate (BHR) to show how many codes can be recovered by the rPPG signal. 
BHR is the percentage of recovered codes for all codes.

\subsection{Attack in User Authentication}

Tab.~\ref{result1} shows the results of a random attack versus an attack using the victim's rPPG signal. 
The success rate of using the victim's rPPG signal is almost doubled compared to random attacks. 
The success rate of the Mean-treated signal is even higher, which is 0.34. 
Although the success rate of rPPG signal is reduced in the Talking and Calculating states, the mean treatment can help mitigate this effect. 
The attack's success rate is 0.35 on average. It is a threat to the authentication model, since real-world authentication systems often allow users to make multiple attempts. 
By analyzing the results, we also found that individual differences were also evident in addition to the quality of the video, which significantly affected the attack's success rate. 
The highest success rate of rPPG signal can reach 0.98 for users. 
In contrast, some users' rPPGs cannot be used for spoofing attacks.
Through the analysis, we found a significant difference in the morphology of some users' rPPG and PPG signals. 

\begin{table}[!ht]
\renewcommand\arraystretch{1.2}
\scriptsize	
\centering
\begin{tabular}{|l| c| c| c| c|c|c|}
\hline
 & \textbf{CHROM}   & \textbf{POS} & \textbf{LGI}& \textbf{PCA}& \textbf{CL\_P}& \textbf{CL\_F} \\\hline
 
MAE (P) & 0.1153   & 0.0762  & 0.0721     & 0.0729   & \textbf{0.0322}  & 0.0592          \\ \hline
RMSE (P) & 0.1564   & 0.1114  & 0.1045   & 0.1051   & \textbf{0.0517}  & 0.0796             \\ \hline
PC (P) & 0.7764    & 0.7985  & 0.8275   & 0.8296    & \textbf{0.9926}  & 0.9925        \\ \hline

\bottomrule

MAE (L) & 0.1572   & 0.1019  & \textbf{0.1001}     & 0.1162   & -  & -          \\ \hline
RMSE (L)  & 0.2008   & 0.1452  & \textbf{0.1419}   & 0.1598   & -  & -            \\ \hline
PC (L)  & \textbf{0.3597}    & 0.3132  & 0.3230   & 0.3212    & - & -        \\ \hline
\bottomrule

MAE (U1) & 0.0771   & \textbf{0.0385}  & 0.0481     & 0.0816   & -  & -          \\ \hline
RMSE (U1) & 0.1207   & \textbf{0.0694}  & 0.0906   & 0.1319   & -  & -            \\ \hline
PC (U1) & 0.6451    & \textbf{0.7380}  & 0.6829   & 0.6099    & - & -        \\ \hline
\bottomrule
MAE (U2) & 0.1319   & 0.0776  & 0.0867     & 0.1131   & \textbf{0.0457}  & 0.0753      \\ \hline
RMSE (U2) & 0.1849   & 0.1332  & 0.1508   & 0.1804   &  \textbf{0.1064}   & 0.1484          \\ \hline
PC (U2) & 0.7554    & 0.8474  & 0.6730   & 0.6018    & \textbf{0.9248}& 0.8787    \\ \hline
\bottomrule
MAE (C) & \textbf{0.1319}   & 0.2271  & 0.2166     & 0.2265   & 0.2167  & 0.2282      \\ \hline
RMSE (C) & \textbf{0.1849}   & 0.2954  & 0.2836   & 0.2957   &  0.4252   & 0.3980          \\ \hline
PC (C) & 0.0110    & 0.0600  & 0.0340   & 0.0060    & 0.9850 & \textbf{0.9880}    \\ \hline

\end{tabular}

\caption{Comparison of the video extracted IPIs with the PPG signal extracted IPIs in different datasets. PC: Pearson correlation coefficient. P: PURE. L: LGI. U1: UBFC\_1. U2: UBFC\_2. C: COHFACE. }
\label{tab:temps}
\end{table}

\begin{table}[!ht]
\centering
\renewcommand\arraystretch{1.2}
\scriptsize	

\begin{tabular}{|l| c| c| c| c|c|c|}
\hline
 & \textbf{CHROM}   & \textbf{POS} & \textbf{LGI}& \textbf{PCA}& \textbf{CL\_P}& \textbf{CL\_F} \\\hline
BHR (P) & 0.5596   & 0.6307  & \textbf{0.6460}     & 0.6302   & 0.6395  & 0.6016      \\ \hline
BHR (L) & 0.5138   & \textbf{0.5756}  & 0.5669     & 0.5638   & -  & -      \\ \hline
BHR (U1) & 0.5544   & 0.6295  & \textbf{0.6460}     & 0.5735   & -  & -    \\ \hline
BHR (U2) & 0.5486   & 0.5810  & 0.5790     & 0.5448   & 0.6344  & \textbf{0.6527}      \\ \hline
BHR (C) & 0.5146   & 0.5205  & 0.5198     & 0.5214   & 0.5743  & \textbf{0.5813}      \\ \hline

\end{tabular}
\caption{The raw video IPI-Trend bit hit rate in different datasets. BHR: Bit hit rate. P: PURE. L: LGI. U1: UBFC\_1. U2: UBFC\_2. C: COHFACE. }
\label{tab:Trend}
\end{table}

\subsection{Attack in IPI-based security protocols}
\label{Attackprotocols}
We report MAE and RMSE for the video extracted IPIs with the PPG signal extracted IPI in Tab.~\ref{tab:temps}. 
The MAE of the recovered IPI in the high-quality original video is below 0.1. 
CL\_P reaches a minimum MAE of 0.03. 
It also recovered an IPI that the Pearson correlation coefficient (PC) is as high as 0.99 with the IPI of PPG. 
It means the IPIs extracted by CL\_P and PPG signals are very similar.

\begin{figure}[!ht]
     \centering
     \begin{subfigure}[b]{\linewidth}
         \centering
         \includegraphics[width=\linewidth]{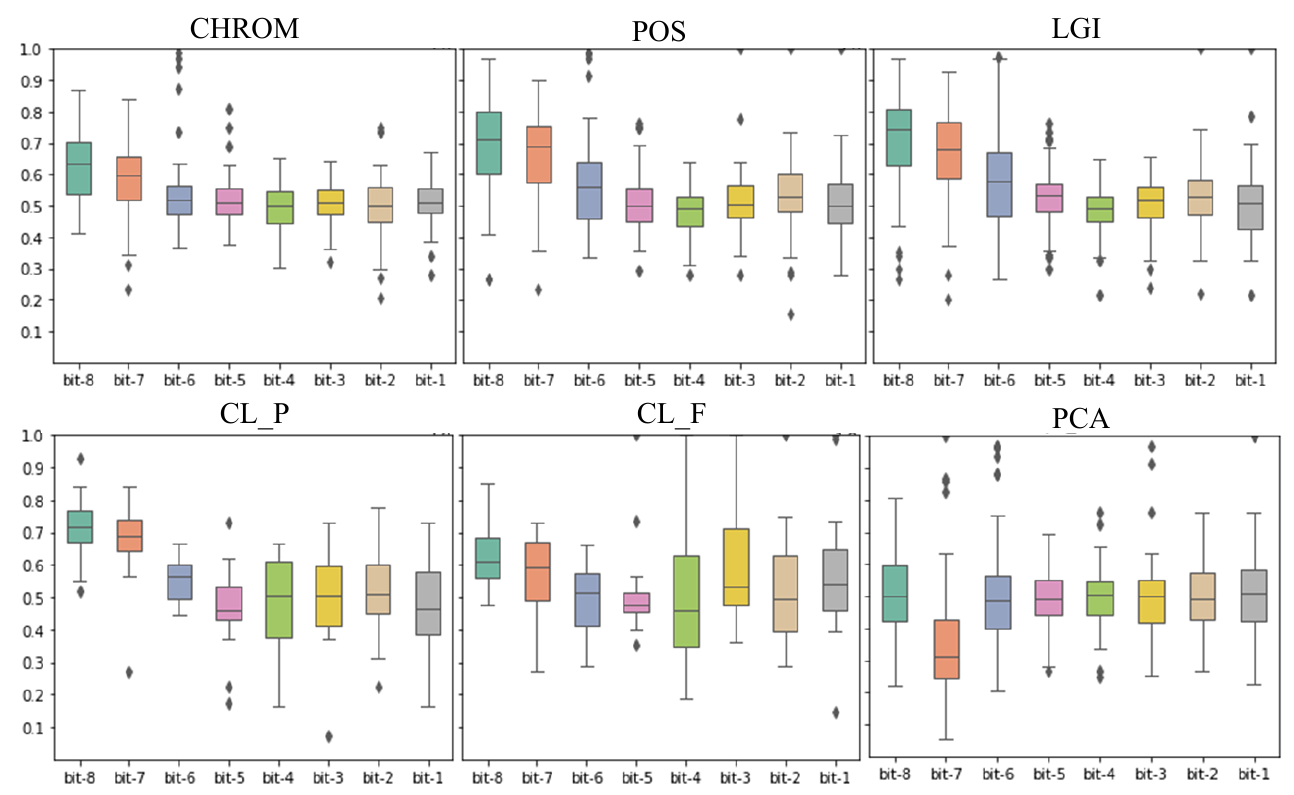}
         \caption{\centering PURE}
         \label{fig:ipicodepure}
     \end{subfigure}
     \hfill
     \begin{subfigure}[b]{\linewidth}
         \centering
         \includegraphics[width=\linewidth]{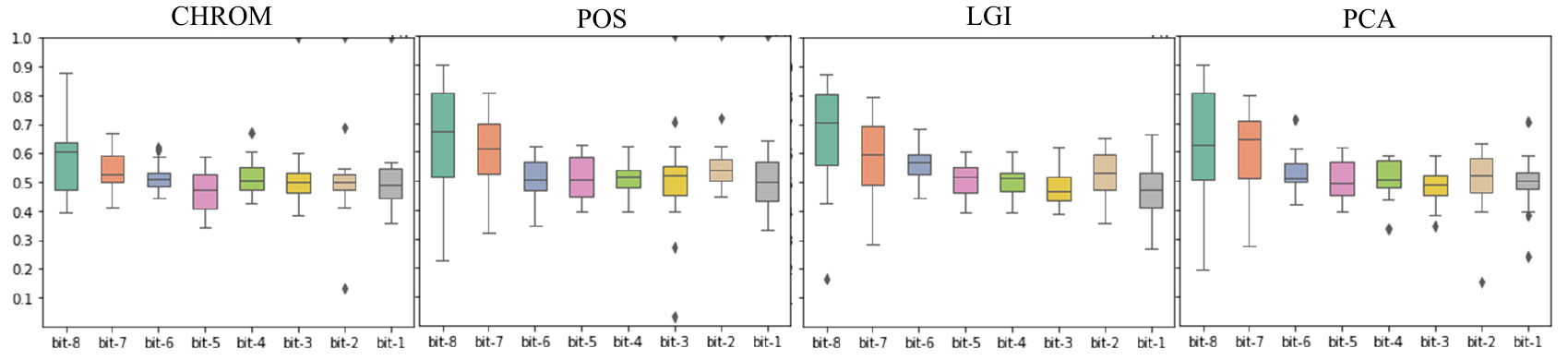}
         \caption{\centering LGI}
         \label{fig:ipicodelgi}
     \end{subfigure}
     \hfill
     \begin{subfigure}[b]{\linewidth}
         \centering
         \includegraphics[width=\linewidth]{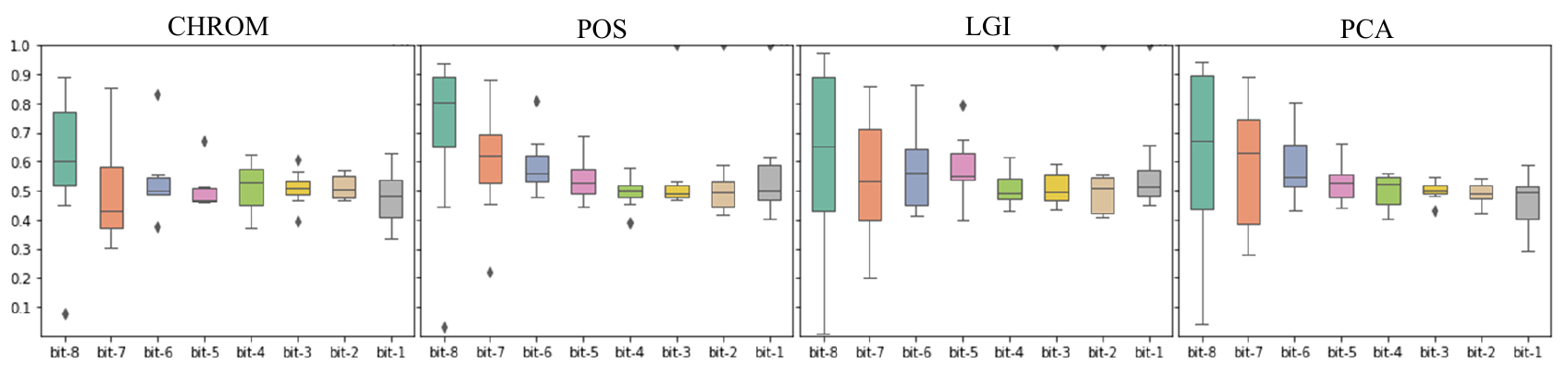}
         \caption{\centering UBFC\_1}
         \label{fig:ipicodeubfc1}
     \end{subfigure}
     \hfill
     \begin{subfigure}[b]{\linewidth}
         \centering
         \includegraphics[width=\linewidth]{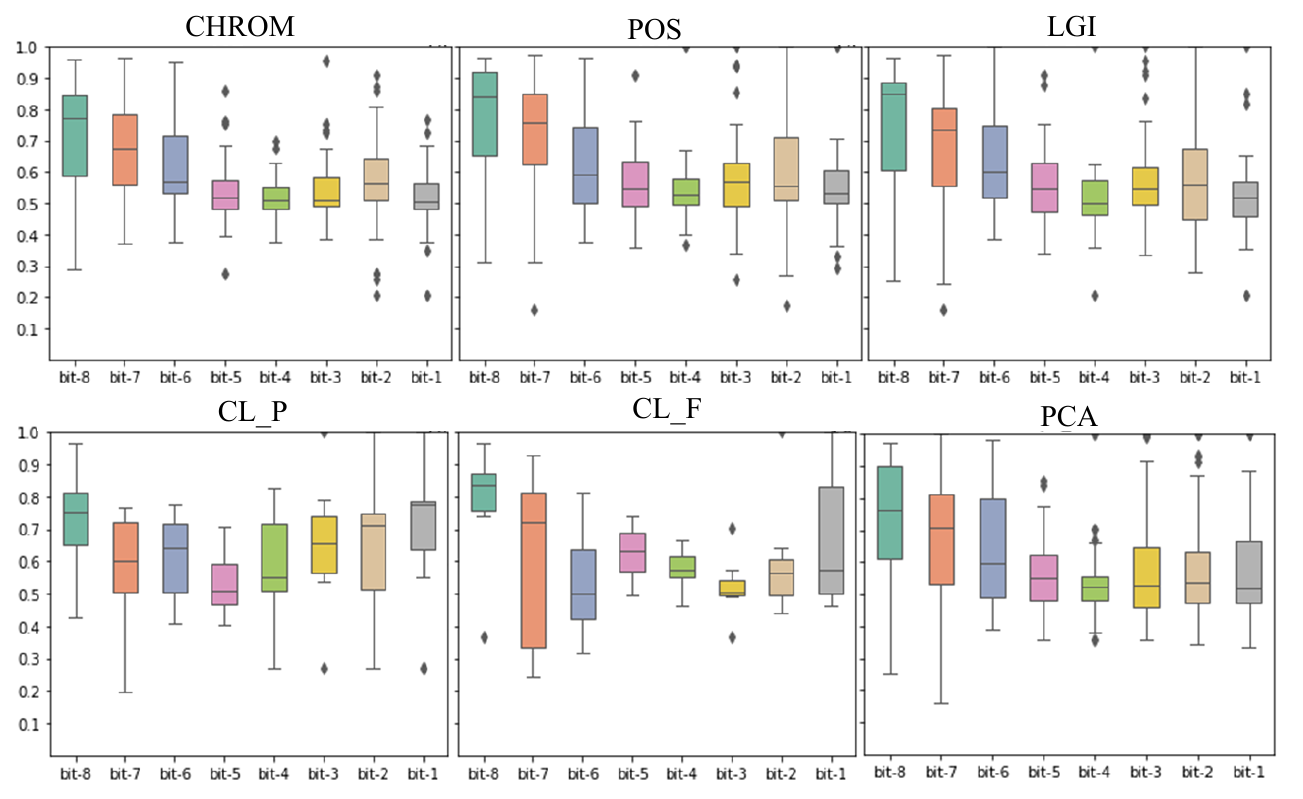}
         \caption{\centering UBFC\_2}
         \label{fig:ipicodeubfc2}
     \end{subfigure}
     
        \caption{The raw video quantile-based IPI coding bit hit rate in different datasets. The horizontal coordinate indicates the bit hit rate.}
        \label{fig:ipicode}
        
\end{figure}

Tab.~\ref{tab:Trend} shows the BHR of the IPI-trend encoding extracted from the original video n different datasets.
The BHR of the original video clip is around 0.6, implying that an attacker has a 60\% success rate of breaking the PPG-based IPI security protocol. In the dataset UBFC\_1, CL\_F reached the highest BHR of 0.65.
Due to the compression, the results in COHEFACE are almost equivalent to a random guess in every rPPG extraction method. Nevertheless, the BHR of CL\_F and CL\_P still reaches 0.58 and 0.57.

For quantile-based IPI coding, the randomness of the bit usually increases as the location of bits decreases. 
Because the performance of the original video extraction from COHFACE was poor in the previous experiments, we choose not to conduct the experiments on COHFACE.
As shown in Fig.~\ref{fig:ipicode}, the BHR of the encoding extracted from the original video varies with the position of bits. 
As the bit position decreases, so does the BHR. 
In the datasets LGI and UBFC\_1, it is obvious to find that the BHR shows a decreasing trend from 0.7 to 0.5 with the decrease of bits.
In UBFC\_2, the CL\_F high BHR even reached above 0.8.
Although the BHR in CL\_P does not vary with bits, their overall BHR is higher than 0.65.

\begin{figure*}[!ht]
  \centering
  \includegraphics[width=\linewidth]{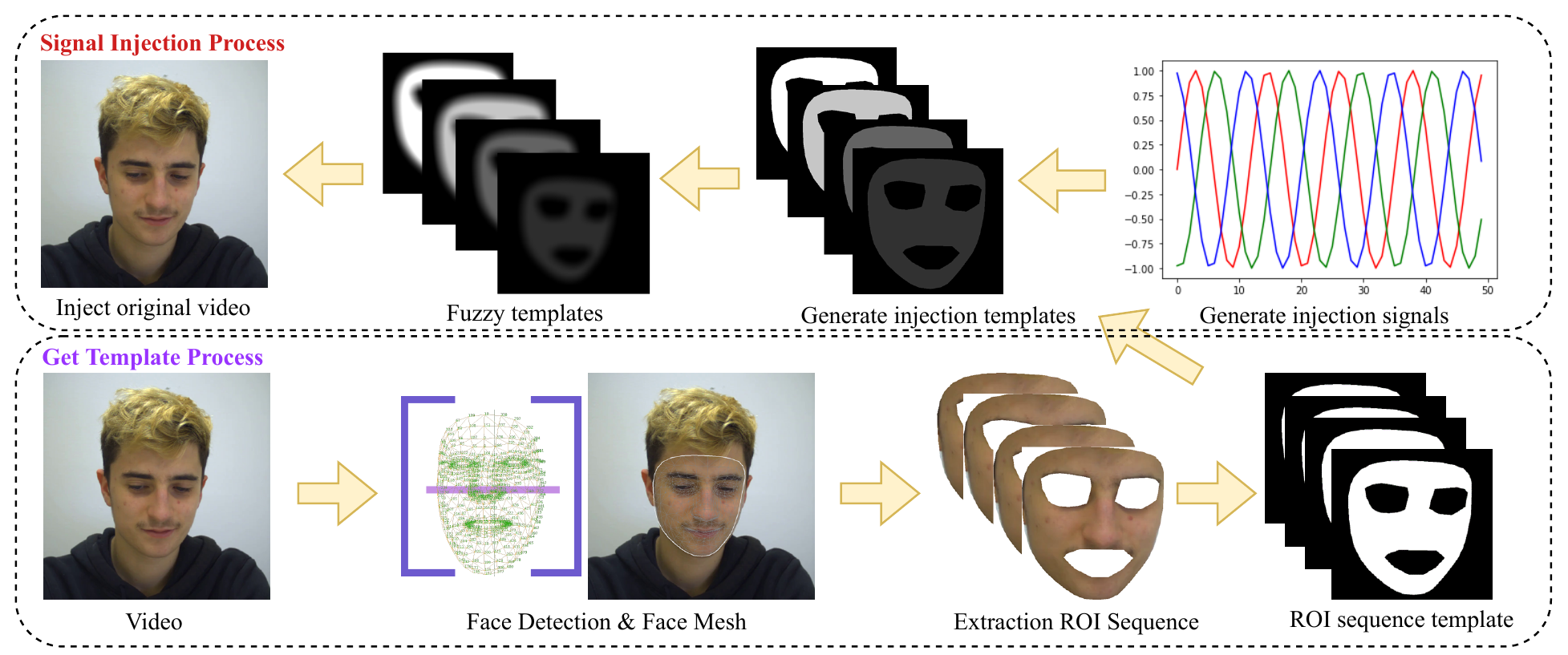}
  \caption{Our proposed workflow for active defence strategy. First, we detect the face area from the original video to distinguish it from the attack strategy. We use a different detection method. Then the face mesh is extracted in the face area. The face mesh includes 468 3D face landmarks. Using face mesh, we remove non-skin regions, such as the eyes and mouth. Next, we use the generated sine signal with the extracted ROI to create a template for injecting the video. Theoretically, we can inject arbitrary waveforms. To make the injected edges imperceptible, we blurred the template. Finally, the template sequence is superimposed on the original video to complete the signal hiding.}
  \label{defence_workflow}
\end{figure*}

\subsection{Discussion}
According to the results in Section \ref{Attackresult}, we conducted an exhaustive analysis on broadly used datasets.
We observed that with the improvement of rPPG extraction methods, state-of-the-art methods have been able to recover the IPI from rPPG signals accurately. 
Both rPPG and PPG signals reflect individual cardiac information, making them contain a great deal of similar information.
While this has facilitated the development of telemedicine, rPPG poses an actual threat to PPG-based biometric authentication, both for user authentication and IPI-based security protocols. 
In user authentication, the spoofing attack achieves a success rate of 35\%, which is a significant improvement of 20\% compared to random attacks. 
Even when the user is in a talking and calculating state, the success rate of rPPG attacks is still much higher than that of random attacks. 
In IPI-based security protocols, the state-of-the-art rPPG method recovered IPI with a BHR above 0.6. 
These results suggest numerous similar morphological features between the rPPG signal and the PPG signal. 
The exposure of these features makes it easier for attackers to compromise PPG signal-based authentication systems.

\section{Defensive Strategies}
\label{defence}

In this section, we investigate passive and active defence strategies against the threats discussed in previous section. 
Passive defence detects between regular and malicious access after an attacker has launched an attack.
Active defence is a strategy taken before an attacker launches an attack. 
We hide the rPPG signal from the face of the video clip, making it difficult for the attacker to use the video to recover the PPG signal. 

\subsubsection{Passive Defence}
Passive defence is primarily used for the scenario of user authentication. 
The input for user authentication is usually the fingertip or wrist PPG signal waveform. 
rPPG signals are collected from the human face, where there are phase differences and morphological differences between rPPG and PPG signals. 
The phase and morphological differences are measured to identify spoofing attacks.

\textbf{Incremental Updating}: When we have an attack sample, we can react to the attack instantly by incrementally updating the model. 
In practice, the attack sample can be obtained by pre-capturing the rPPG signal from the user's face. 
However, we may not be able to simulate the attack sample in this way accurately.

\textbf{Anomaly Detection}: It is also known as outlier analysis. 
It only needs to learn the user's samples before detecting malicious samples by excluding outliers. 
This paper uses Isolation Forest and One-Class SVM as anomaly detectors.

Passive defence usually follows the successful attack. Both incremental updates and anomaly detection can potentially affect the performance of the original user authentication model.

\subsubsection{Active Defence}
To solve the limitation of passive defence, we propose an active defence strategy, namely SigH. As shown in Fig.~\ref{defence_workflow},
we hide the exposed rPPG signal by injecting the noise signal into the original video.
It allows us to start our defence before being attacked, minimising the impact on the original model. 
Our active defence strategy consists of the following components:

\textbf{Extraction of ROI area}: First we use BlazeFace \cite{bazarevsky2019blazeface} to detect the face area. 
It can detect faces quickly (200 to 1000 FPS) on a mobile GPU. 
Then we apply the method in \cite{kartynnik2019real} to get the face landmarks. 
It enables real-time (100-1000 FPS) inference of individual camera inputs on mobile devices. 
After that, we will get 468 3D face landmarks. These landmarks mark the locations of feature points in the face. 
For example, the tip of the nose and the top of the left/right cheekbones.

\begin{table*}[!ht]
\renewcommand\arraystretch{1.2}
\centering
\setlength{\tabcolsep}{5mm}{\begin{tabular}{|l| c| c| c| c| c|  }
\hline

\multirow{2}{*}{ }  & \multicolumn{3}{|c|}{\textbf{Performance of the authentication model}} & \multicolumn{2}{|c|}{\textbf{Spoofing attack success rate}} \\ \cline{2-6}

 & \textbf{F1-Score}  & \textbf{Precision} & \textbf{Recall} & \textbf{rPPG} & \textbf{Mean-rPPG} \\\hline
Original Model & \textbf{0.8735}   & \textbf{0.9127}   & \textbf{0.8548} &  0.2517  & 0.3427 \\ \hline
Incremental Updating & 0.8269   & 0.8166 & 0.8645 & 0.0067 & 0.0275 \\ \hline
OneClassSVM  & 0.7369   & 0.8141  & 0.7138  &  0.0420 &  0.0572 \\ \hline
IsolationForest & 0.7364   & 0.8265   & 0.7079  & 0.0430 & 0.0586  \\ \hline
\end{tabular}}

\caption{Performance of three passive defence strategies (Incremental Updating, One-Class SVM, and Isolation Forest). The left half shows the impact on the performance of the original model. The right half is the success rate of the victim rPPG signal attack authentication model.}
\label{result2}
\end{table*}

\textbf{Create Injection Template}: For each frame $C_(t)$ in the video, We use the facial region as the ROI and exclude the eye and mouth regions. 
As shown in the formula below, in the template, the RGB value of the ROI area is set to 1 and the rest of the area is set to 0.
\begin{equation}
C_{temp}(t) = \left\{\begin{matrix}
 C_{ROI}(t) = 1 &  \\
 C_{O}(t) = 0 &  \\
\end{matrix}\right. 
\end{equation}
The sine signal is superimposed on the template. 
\begin{equation}
Kernel = \frac{1}{30}\begin{bmatrix}
 1& \dots &1  \\
 \dots & \dots & \dots \\
 1& \dots & 1 \\
\end{bmatrix}_{30\times30}
\end{equation}
The template is fuzzed with a convolution kernel. 
\begin{equation}
C_{final\_temp}(t) = C_{temp}(t) \times Sine(t) * Kernel
\end{equation}

\begin{figure}[!ht]
  \centering
  \includegraphics[width=\linewidth]{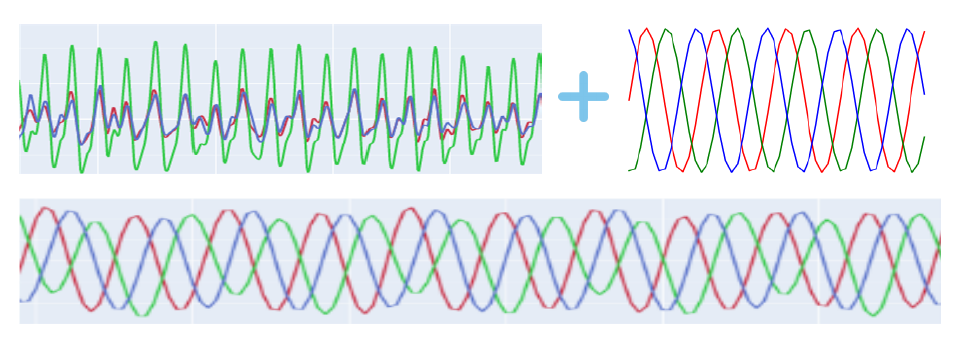}
  \caption{An active defence example. The RGB sequence value extracted directly from the original video is in the upper left corner. The upper right corner is the injected signal. Below is the RGB sequence value extracted from the processed video.}
  \label{activewave}
\end{figure}

Finally, we superimpose the template sequence on the original video frames to complete the signal hiding. 
$Frame_f(t)$ represents the $t$-th frame of the processed video. $Frame_o(t)$ indicates the $t$-th frame of the original video clip.
\begin{equation}
Frame_f(t) =  Frame_o(t) + C_{final\_temp}(t)
\end{equation}

Fig.~\ref{activewave} shows the results of our signal injection in the video. We observe that the injected signal is completely different from the original signal.

\begin{table}[!t]
\renewcommand\arraystretch{1.2}
\centering
\setlength{\tabcolsep}{3mm}{\begin{tabular}{|l| c| c| }
\hline
 Video Type & \textbf{CHROM-rPPG}  & \textbf{Mean-rPPG}   \\\hline
Original Video & 0.7005   & 1.0000  \\ \hline
SigH Video  & 0.0291   & 0.0577  \\ \hline
\end{tabular}}
\caption{ Performance of active defence strategy in user authentication. SigH: Our proposed signal hiding method. }
\label{result3}
\end{table}

\section{Defence Evaluation}
\label{Defenceresult}
We evaluate the performance of our defensive strategies (see Section~\ref{defence}) in this section. The same datasets as in the previous section are used in this section. 
We use MediaPipe\footnote{\url{https://google.github.io/mediapipe/}} to obtain the face mesh for each video frame. 
Then the ROI template is extracted from the face mesh using OpenCV\footnote{\url{https://opencv.org/}} before we perform signal injection to the templates.

\subsection{Defence in User Authentication}

Tab.~\ref{result2} shows the performance of the passive defence strategy. 
These results show that all passive defence strategies significantly reduce the attack's success rate. 
The success rate of rPPG attacks has been reduced from 0.25 to at least 0.04. 
Mean-treated rPPG was reduced to 0.05. 
However, we observe that the passive defence strategy seriously affects the performance of the authentication model system. 
The incremental updating method has the least impact on the authentication system. 
The F1-Score of the model dropped from 0.87 to 0.82. 
It reduces the success rate of rPPG attacks and makes the model less generalized.
IsolationForest and OneClassSVM have a significant impact. 
The F1-Score of the model is around 0.73. 
Unfortunately, it propagates the error rate to the authentication model while blocking the attack.

We propose an active defense strategy to solve the problems in passive defence. 
To maximize the active defence strategy's performance, we selected 13 video clips from the original videos with the success rate of being attacked greater than 0.5. 
As shown in Tab.~\ref{result3}, the attack's success rate using the rPPG signal extracted from the original video reaches 0.7, and the mean-treated rPPG signal even reaches 1. 
The success rate of video rPPG attacks processed by the active defence strategy is reduced to 0.02, and the mean-treated rPPG is decreased to 0.05, almost achieving the same performance as the passive defence. 
Moreover, the active defence strategy does not affect the original model's performance.

\begin{table}[!t]
\renewcommand\arraystretch{1.2}
\scriptsize	
\centering
\begin{tabular}{|l| c| c| c| c|c|c|}
\hline
 & \textbf{CHROM}   & \textbf{POS} & \textbf{LGI}& \textbf{PCA}& \textbf{CL\_P}& \textbf{CL\_F} \\\hline
 
MAE-S (P) & 0.1729   & 0.2087  & \textbf{0.1661}     & 0.1780   & 0.2039  & 0.1691          \\ \hline
RMSE-S (P) & 0.2222   & 0.2536  & 0.2153   & 0.2279   & 0.2422  & \textbf{0.2086}             \\ \hline
PC-S (P) & -0.046    & -0.037 & \textbf{-0.057}   & -0.055    & 0.2897  & 0.3917        \\ \hline

\bottomrule

MAE-S (L) & 0.2782   & \textbf{0.2676}  & 0.2872     & 0.2899   & -  & -          \\ \hline
RMSE-S (L)  & 0.3419   & \textbf{0.3274}  & 0.3466   & 0.3509   & -  & -            \\ \hline
PC-S (L)  & 0.0317    & 0.0112  & \textbf{-0.009}   & -0.005    & - & -        \\ \hline
\bottomrule

MAE-S (U1) & 0.3141   & 0.2481  & \textbf{0.2472}     & 0.3417   & -  & -          \\ \hline
RMSE-S (U1) & 0.3687   & 0.3061  & \textbf{0.3051}   & 0.3893   & -  & -            \\ \hline
PC-S (U1) & 0.0003    & 0.0115  & 0.0098   & \textbf{-0.050}    & - & -        \\ \hline
\bottomrule
MAE-S (U2) & 0.3597   & 0.3063  & 0.3170     & 0.4012   & 0.3610  & \textbf{0.2728}      \\ \hline
RMSE-S (U2) & 0.4077   & 0.3574  & 0.3689   & 0.4403   &  0.4045   & \textbf{0.3200}          \\ \hline
PC-S (U2) & 0.0397    & \textbf{0.0058}  & 0.0163   & 0.0447    & 0.4759& 0.7647    \\ \hline
\bottomrule
MAE-S (C) & \textbf{0.2884}   & 0.2948  & 0.2943     & 0.2942   & 0.4112  & 0.4629      \\ \hline
RMSE-S (C) & \textbf{0.3209}   & 0.3242  & 0.3237   & 0.3234   &  0.4873   & 0.5359          \\ \hline
PC-S (C) & -0.007    & \textbf{-0.033}  & -0.025   & -0.023    & 0.3360 & 0.2036    \\ \hline

\end{tabular}

\caption{Comparison of the video extracted IPIs with the PPG signal extracted IPIs in different datasets. PC: Pearson correlation coefficient. P: PURE. L: LGI. U1: UBFC\_1. U2: UBFC\_2. C: COHFACE. -S: Signal acquired from SigH processed video.}
\label{tab:temps1}
\end{table}

\begin{figure}[!ht]
     \centering
     \begin{subfigure}[b]{\linewidth}
         \centering
         \includegraphics[width=\linewidth]{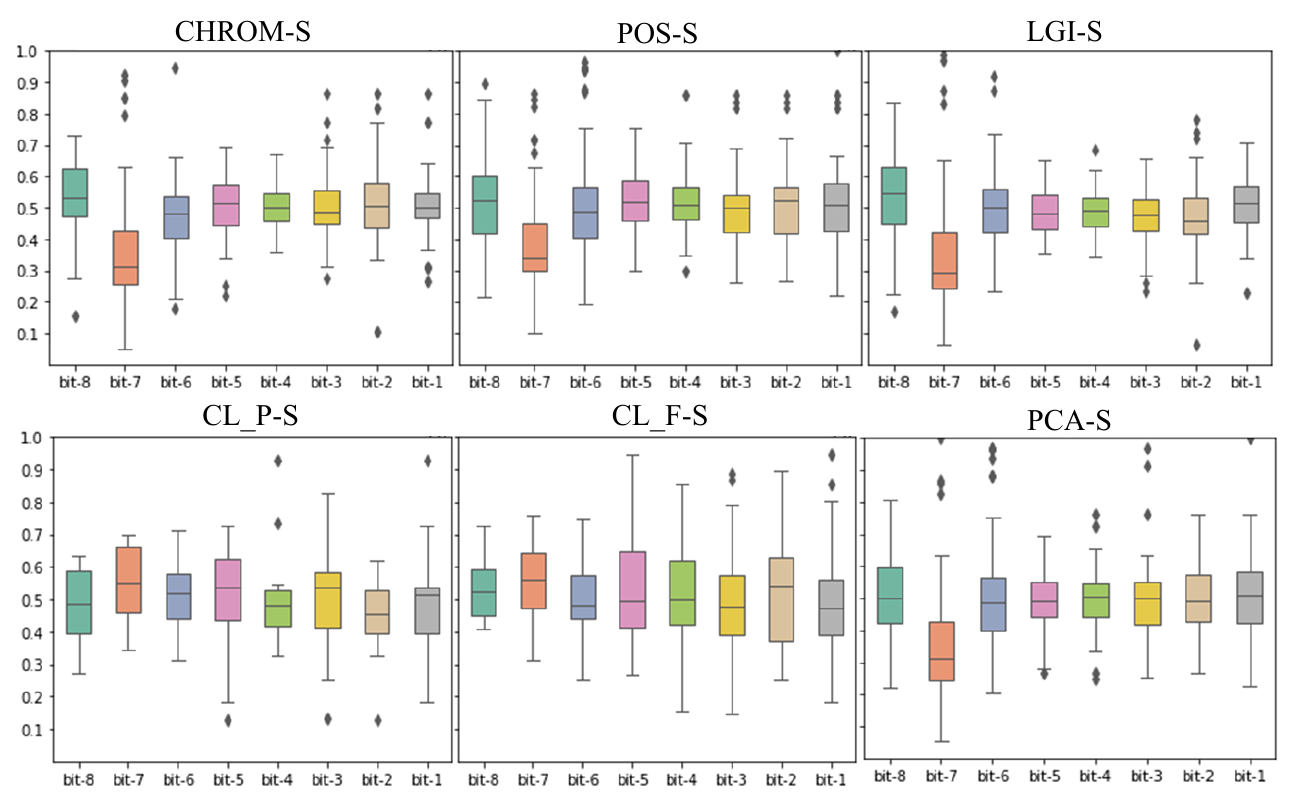}
         \caption{\centering PURE}
         \label{fig:ipicodepureh}
     \end{subfigure}
     \hfill
     \begin{subfigure}[b]{\linewidth}
         \centering
         \includegraphics[width=\linewidth]{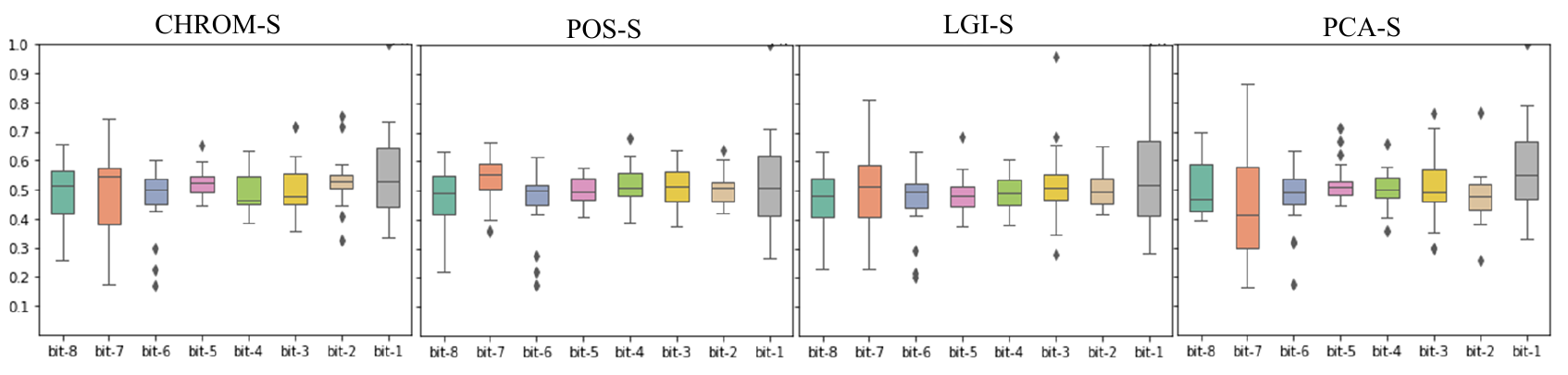}
         \caption{\centering LGI}
         \label{fig:ipicodelgih}
     \end{subfigure}
     \hfill
     \begin{subfigure}[b]{\linewidth}
         \centering
         \includegraphics[width=\linewidth]{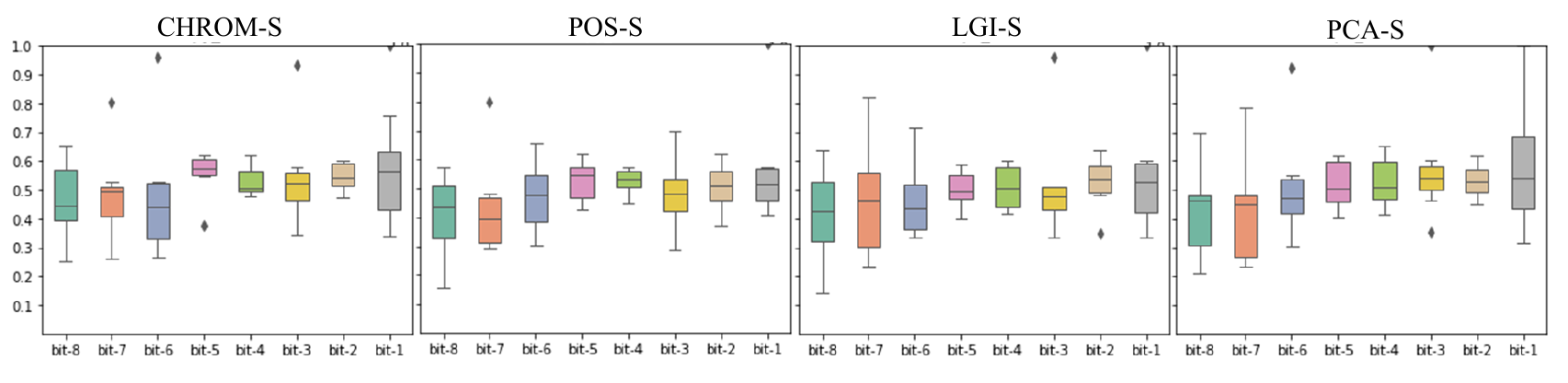}
         \caption{\centering UBFC\_1}
         \label{fig:ipicodeubfc1h}
     \end{subfigure}
     \hfill
     \begin{subfigure}[b]{\linewidth}
         \centering
         \includegraphics[width=\linewidth]{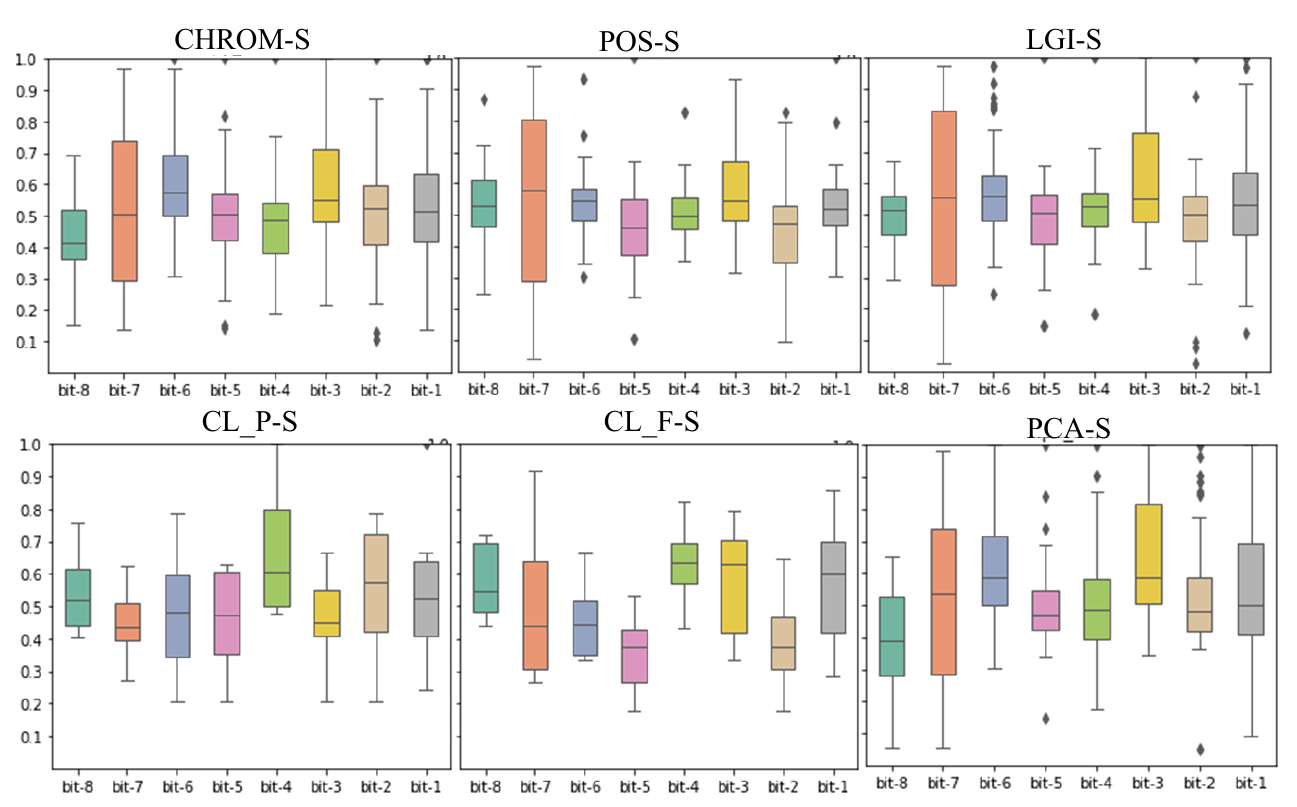}
         \caption{\centering UBFC\_2}
         \label{fig:ipicodeubfc2h}
     \end{subfigure}
     
        \caption{The processed video quantile-based IPI coding bit hit rate in different datasets. The horizontal coordinate indicates the bit hit rate. -S: Signal acquired from SigH processed video.}
        \label{fig:ipicodeh}
        
\end{figure}

\begin{table}[!t]
\centering
\renewcommand\arraystretch{1.2}
\scriptsize	

\begin{tabular}{|l| c| c| c| c|c|c|}
\hline
 & \textbf{CHROM}   & \textbf{POS} & \textbf{LGI}& \textbf{PCA}& \textbf{CL\_P}& \textbf{CL\_F} \\\hline
BHR-S (P) & 0.5516   & 0.5923  & 0.5557     & 0.5758   & 0.5410  & \textbf{0.5935}      \\ \hline
BHR-S (L) & \textbf{0.5412}   & 0.5208  & 0.5179     & 0.5362   & -  & -      \\ \hline
BHR-S (U1) & \textbf{0.6042}   & 0.5210  & 0.5374     & 0.5700   & -  & -    \\ \hline
BHR-S (U2) & 0.5735   & 0.5563  & 0.5406     & \textbf{0.5787}   & 0.5709  & 0.5283      \\ \hline
BHR-S (C) & 0.5388   & \textbf{0.5432}  & 0.5397     & 0.5396   & 0.5312  & 0.5259      \\ \hline

\end{tabular}
\caption{The processed video IPI-trend bit hit rate in different datasets. BHR: Bit hit rate. P: PURE. L: LGI. U1: UBFC\_1. U2: UBFC\_2. C: COHFACE.  -S: Signal acquired from SigH processed video.}
\label{tab:Trend1}
\end{table}

\subsection{Defence in IPI-based security protocols}
Compared to the best results (CL\_P, dataset PURE) of the original video extracted IPI,  in our processed video, MAE increased to 0.2, PC was reduced to 0.28. 
In the other datasets, MSE increased to some extent. 
However, we found that the variation was minimal in COHFACE because the video clips in COHFACE are compressed. 
The MAE of the IPI recovered from COHFACE has reached 0.2s. 
Their PC is around 0.05 for all methods except CL\_P and CL\_F. 
It indicates that the IPI recovered by other methods has a low correlation with the IPI of PPG. 
After processing the video, even the state-of-the-art method PC is reduced to approximately 0.3.

Contrary to the result in Section~\ref{Attackprotocols}, Tab.~\ref{tab:Trend1} shows that the BHR of videos processed by active defense methods is approximately 0.5, which is almost equal to random guessing. 
As shown in Fig.~\ref{fig:ipicodeh}, in the processed video, BHR remains steady at approximately 0.5 with the bit position. 
For example, as shown in Fig.~\ref{fig:ipicodepure}, the LGI and POS methods gradually decrease in BHR from 0.7 to 0.5 as the bits go down. 
In contrast, as shown in Fig.~\ref{fig:ipicodepureh}, BHR does not change with the bits.
It means that our proposed active defense method completely hides the PPG signal of the user in the video.

\subsection{Discussion}
 
A common strategy to combat spoofing attacks is by introducing detection components.
Nevertheless, the detection component is usually located in front of the recognition pipeline, and its errors will be propagated to the recognition model, affecting the overall model recognition performance.
In particular, when the spoofed signal is similar to a real signal, the model's false-negative rate increases significantly.
We observed that low-quality videos lower the success of spoofing attacks. 
For example, in the COHFACE dataset, the quality of the obtained rPPG signals is relatively poor, and the success rate of the attack is lower than in other datasets.
Compared to other datasets, the video clips in COHFACE have a frame rate of 20 FPS, $640\times 480$ resolution, and 255 Kbps bit rate.
However, preventing attacks by sacrificing video quality is not always practical. 
Thus, we propose an active defence to prevent leaking the target's PPG signals by modifying the RGB pixel values of the facial skin in the video frame.
Video hosting platforms like YouTube and TikTok can successfully prevent signal leakage by batch processing users' videos.

\section{Conclusion}
\label{conclusion}
Recently, emerging biometric solutions using physiological signals have gained widespread attention. 
In particular, the PPG signal is easy to collect and unobservable to remote attackers. 
However, the rPPG signal breaks this unobservability.
To comprehensively analyze the impact of the rPPG signal, we conducted experiments on five datasets (PURE, UBFC\_rPPG, UBFC\_Phys, LGI\_PPGI, and COHFACE). 
We found that user authentication and IPI-based security protocols are vulnerable to rPPG signal spoofing attacks.
To mitigate this spoofing attack, we propose an active defence scheme. 
It has the most negligible impact on the performance of the original model compared to the passive defence. 
Before releasing HD video, we recommend video platforms using active defence strategies in bulk to mitigate rPPG signal leakage.



%





\ifCLASSOPTIONcaptionsoff
  \newpage
\fi





\bibliographystyle{IEEEtran}
\bibliography{IEEEabrv,Bibliography}

\vfill


\end{document}